\documentclass[fleqn,usenatbib]{mnras}

\usepackage{newtxtext,newtxmath}

\usepackage[T1]{fontenc}
\usepackage{ae,aecompl}

\usepackage{graphicx}	
\usepackage{amsmath}	
\usepackage{amssymb}	
\usepackage{multirow}

\title[NGC\,6876 and its globular cluster system]{Early-type galaxies in low-density environments: NGC\,6876 explored through its globular cluster system}

\author[A. I. Ennis et al.]{
Ana I. Ennis,$^{1,2}$\thanks{E-mail: anaennis@fcaglp.unlp.edu.ar}
Lilia P. Bassino$^{1,2}$,
Juan P. Caso$^{1,2}$ and
Bruno J. De B\'ortoli$^{1,2}$
\\
$^{1}$Facultad de Ciencias Astron\'omicas y Geof\'isicas de la Universidad Nacional de La Plata, and Instituto de Astrof\'isica de La Plata \\ (CCT La Plata -- CONICET, UNLP), Paseo del Bosque S/N, B1900FWA La Plata, Argentina\\   
$^{2}$Consejo Nacional de Investigaciones Cient\'ificas y T\'ecnicas, Rivadavia 1917, C1033AAJ Ciudad Aut\'onoma de Buenos Aires, Argentina}

\date{Accepted XXX. Received YYY; in original form ZZZ}

\pubyear{2018}

\begin{document}
\label{firstpage}
\pagerange{\pageref{firstpage}--\pageref{lastpage}}
\maketitle

\begin{abstract}
We present the results of a photometric study of the early-type galaxy NGC\,6876 and the surrounding globular cluster system (GCS). The host galaxy is a massive elliptical, the brightest of this type in the Pavo Group. According to its intrinsic brightness ($M_{\rm v} \sim$ -22.7), it is expected to belong to a galaxy cluster instead of a poor group. Observational material consists of $g'$, $r'$, $i'$ images obtained with the Gemini/GMOS camera. The selected globular cluster (GC) candidates present a clear bimodal colour distribution at different galactocentric radii, with mean colours and dispersions for the metal-poor (``blue'') and metal-rich (``red'') typical of old GCs. The red subpopulation dominates close to the galaxy centre, in addition to the radial projected distribution showing that they are more concentrated towards the galaxy centre. The azimuthal projected distribution shows an overdensity in the red subpopulation in the direction of a trail observed in X-ray that could be evidence of interactions with its spiral neighbour NGC\,6872. The turn-over of the luminosity function gives an estimated distance modulus $(m - M) \approx$ 33.5 and the total population amounts to 9400\,GCs, i.e. a quite populous system. The halo mass obtained using the number ratio (i.e. the number of GCs with respect to the baryonic and dark mass) gives a total of $\sim10^{13}$, meaning it is a very massive galaxy, given the environment.

\end{abstract}




\section{Introduction}

Environment plays a major role in determining several features of galaxies, such as their Hubble type, colour and star formation rate \citep{zhang2013}, as well as in conditioning the evolutionary processes they undergo. Early-type galaxies (ETGs) are usually found in high-density environments, which is why most studies focus on galaxies inhabiting clusters. However, in low-density environments accretion mechanisms can be less efficient than in clusters \citep{cho2012}, and allow us to study growth and evolutionary processes in action \citep[e.g.][]{bassino2017}.

Taking the environment into account proves to be key to understand galaxy evolution, since statistical studies \citep{tal2009} have shown that it is in massive ETGs inhabiting low-density environments that tidal features are most frequently found, meaning it is very common for them to be experiencing gravitational interactions. \cite{hirschmann2013} finds through studying simulated field ETGs that a large fraction of them suffers late mergers, in which there is usually very little gas involved \citep{jimenez2011}. These interactions or ``dry mergers'' could explain how the halos of ETGs in low-density environments keep on growing while maintaining the average age and metallicity of their population \citep{thomas,lacerna2016}. The fact that processes of this sort are expected in galaxies in these environments has caused an increase in their study \citep{salinas2015,caso2017,escudero2018}.

The analysis of objects that are old enough to have belonged to their host galaxy since its formation, and that have survived ever since, is remarkably useful for untangling all the processes the galaxy has gone through since formation. In this sense, globular clusters (GCs) are perfect candidates, 
since their estimated ages are $> 10 Gyr$ \citep{peng2006}. They also offer the advantage of being compact and intrinsically bright enough that they can be detected as point-like sources in galaxies as far $\approx 200\,$Mpc \citep[e.g.][]{harris2014}.  

One of the most studied characteristics of globular cluster systems (GCS) is their colour distribution, which is very often found to be bimodal in massive ETGs, indicating the presence of two subpopulations \citep[e.g.][]{harris2015}. Though this could be a result of either a difference in age or in metallicity, spectroscopic studies indicate that there is no age difference among (old) GCs to explain such a gap \citep[e.g.][]{woodley2010}. Colour bimodality is then understood as a difference in metallicity, with blue GCs having significantly less metal content than red ones \citep[e.g.][]{brodie2012}. In addition to the separation in colour, these GC subpopulations show other differences, such as their kinematics and spatial distribution \citep[e.g.][]{schuberth2010,pota2015,caso2017}. The origin of these subpopulations has not been fully explained yet. 

 There are a few mechanisms that do reproduce the bimodality. The hierarchic merger \citep{beasley2002} is based on semi-analitical model of formation, combining elements from older models. Blue GCs are formed in the early Universe (z>5) until a process, most likely cosmic reionization \citep{vandenbergh2001}, puts a stop to it. Afterwards, gas-rich major mergers trigger the formation of the red GCs. However, the more accepted models nowadays are the assembly scenario \citep{tonini2013} and the merging scenario suggested by \cite{li2014}. Both use results from the Millenium simulation, the first one being based on the fusion rates obtained from it, and proposing that the red subpopulation is formed in the host galaxy at around $z\approx 2$ whereas the blue one is accreted from satellite galaxies. The latter uses the halo merger trees obtained from the Millenium simulations and an empirical relation between the galaxy's mass and metallicity. In this model, the bimodality emerges because of red GCs forming in mergers of massive halos, while blue GCs were formed in previous mergers, with less massive halos involved.
 
Recent works \citep{pfeffer2018,kruijssen2018} combined the EAGLE model for galaxy formation simulations \citep{schaye2015,crain2015} with the MOSAICS model for cluster formation and evolution \citep{kruijssen2011} in order to study the connection between the evolution of GCS and their host galaxy. In \cite{kruijssen2018}, the analysis of the Milky Way (MW) and its cluster population showed that at least 40\% of the clusters had formed \textit{ex situ}, identifying their progenitors as massive galaxies accreted by the MW early on its history.

Recently, a third subpopulation has been found in several galaxies. All of them show signs of interactions which had been previously detected, and stellar population models fitted to this third subpopulation estimated an age for it that matches the time past since these merger or accretion events. This hints at the possibility that this would be the result of star formation events being triggered by interactions, which makes this third subpopulation a subject of interest since they would be globular clusters currently at a much earlier stage of formation than the ones usually observed \citep{caso2015,bassino2017,sesto2018}.

\vspace{0.4cm}

\begin{figure}
 \includegraphics[width=\columnwidth]{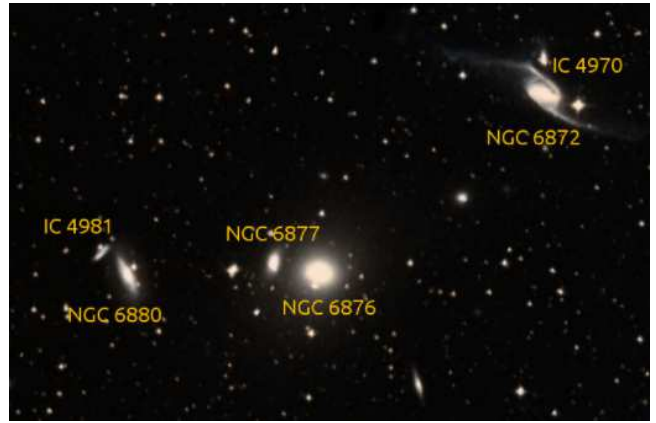}
 \caption{DSS image of the group centered on NGC\,6876. North is up, East is to the left.}
 \label{fig:cielo}
\end{figure}

NGC\,6876 is a massive elliptical galaxy located in the Pavo galaxy group (Figure\,\ref{fig:cielo}). The different values for the distance modulus of NGC\,6876 found in the NASA/IPAC Extragalactic Database (NED) present a wide dispersion. In this work, we adopt the mean of the two corrected values from \citet{blakeslee2001}, based on the fundamental plane and on the SBF method respectively, obtaining a distance of $ \sim45$\,Mpc. 

From the SIMBAD database, its apparent blue magnitude is $B=11.76$, and considering the median colors for elliptical galaxies given by \cite{fukugita1996} and the absorption maps constructed by \cite{schlafly11}, it translates into an estimated absolute visual magnitude of $M_{V}\approx-22.7$. 

Though ETGs as luminous as NGC\,6876 are usually found in rich clusters, the Pavo galaxy group consists of barely thirteen galaxies, confirmed through radial velocities, with a velocity dispersion of 425 ${\rm km}\,{\rm s}^{-1}$ \cite[][and references therein]{machacek2009}. In addition, it appears to be a dynamically young group. Many members have less massive companions, showing strong signs of interaction between them. NGC\,6876 in particular has an elliptical companion, NGC\,6877, at $1.4\,\rm{arcmin}$ to the east with a difference in radial velocity of $\Delta V_{r}\approx300 \,km \,s^{-1}$, which could be in an early pre-merger phase \cite{machacek2009}. 

However, the interesting suggestion that NGC\,6876 might be interacting with other members is not related to these less massive neighbors. Instead, \textit{XMM-Newton} observations revealed a $\approx$\,90\,kpc trail of X-ray emission hotter ($\approx 1$\,KeV) than the Pavo intergalactic medium (IGM) that links NGC\,6876 to its nearest massive neighbor, NGC\,6872 \citep{machacek2005} which is located at a projected distance of $8.7$\,arcmin, $\sim113$\,kpc given our assumed distance.  The latter is a spiral galaxy which is currently interacting with its companion, IC\,4970 \citep{mihos1993,machacek2008}. Before the discovery of this trail, NGC\,6876 was not believed to have any relation with the morphological distortions of NGC\,6872, since there is a significant difference in their radial velocities ($\sim$\,13\%, NED), and their projected distance is eight times larger than the projected distance between NGC\,6872 and IC\,4970.
Besides being the first hint that these two massive galaxies might be interacting, this trail is important since there are very few trails associated to large spiral galaxies, and it is the first one observed in a low-density group \citep{machacek2005}.

\cite{machacek2005} concludes, from analyzing the \textit{XMM-Newton} observations of the X-ray trail, that it cannot be entirely made of tidally stripped gas from NGC\,6876. Tidal interactions strong enough to have removed that amount of hot gas should have also altered the stellar distribution, and such distortions have not been found yet. Nonetheless, interactions cannot be ruled out, though it is also suggested that the trail is primarily associated with NGC\,6872, since its surface brightness increases near the spiral galaxy. In a further analysis of the trail, \citet{horellou2007} find no signs of interaction with NGC\,6872 in the HI distribution of NGC\,6876. Though their simulations do not model the motions of NGC\,6872 through the IGM or the gas environment, they suggest the trail might be a result of Bondi-Hoyle accretion around NGC\,6872, where the IGM gas of the group suffers from gravitational focus because of the galaxy passing through the core of the group, resulting in its adiabatic heating \citep{bondi1944}.

For better resolution, \cite{machacek2009} combines \textit{Spitzer}'s mid-infrared and \textit{Chandra} X-ray observations of NGC\,6872 and NGC\,6876 with archival optical as well as HI data. The emission excess at $8\rm{\mu}m$ found in NGC\,6876 is weak and mostly stellar. The non-stellar portion appears to come from leakage from emission produced by silicate dust grains ejected from the atmospheres of evolved AGB stars. 

A more recent study of NGC\,6872 \citep{eufrasio2014}, however, goes back to the original assumption that these two galaxies might have interacted not that long ago. Using UV-to-IR archival data, this analysis of the spiral galaxy reveals NGC\,6876 might have shaped its neighbor over a long period of time, contributing to the bar before the interactions between NGC\,6872 and its minor companion had started. In this scenario, it is reasonable to expect to find some evidence of this interaction in NGC\,6876 as well.

In this work, we perform a photometric study of the globular cluster system of NGC\,6876, based on Gemini observations. We analyze the colour distribution and,  after separating the GC subpopulations, we present their individual spatial distributions, as well as the luminosity function of the entire system. In addition, we study the surface-brightness distribution of the galaxy. Preliminary results \citep{ennis2017a,ennis2017b} have shown a clear bimodality in the colour distribution, with the two GC subpopulations following the same spatial distributions as typically found in the literature. 

The structure of this paper is as follows. In Section\,2, the observations and reduction process are described. In Section\,3, we present the GC selection and in Section\,4 the analysis of the globular cluster system. In Section\,5 we study the host galaxy surface brightness distribution, while a discussion of our findings is presented in Section\,6. Finally, in Section\,7 we report our conclusions.
 
\section{Observations and Data Reduction}


The data set was taken over the course of several nights spread throughout August, September and November of 2013, with the Gemini Multi-Object Spectrograph (GMOS) camera, mounted on the Gemini South telescope (programme GS-2013B-Q-37, PI: J.P. Caso). 
 
A single field containing the galaxy was observed in $g'$, $r'$, and $i'$ filters, as detailed in Table \ref{tab:exp}, with $2 \times 2$ binning which resulted in a scale of $0.146\,{\rm arcsec}\,{\rm pixel}^{-1}$.

\begin{table}
\centering
 \caption{Exposure times and central wavelengths for each filter}
 \label{tab:exp}
 \begin{tabular}{|l|l|l|}
  \hline
  Filter & $\lambda_{\rm eff}$ [nm] & Exposure Time [s] \\
  \hline
  $g'$ & 475 & $8\times800$\\[2pt] 
  $r'$ & 630 & $6\times450$\\[2pt]
  $i'$ & 780 & $9\times500$\\[2pt]
  \hline
  	\hline
    
\end{tabular}
 
\end{table}

Bias and flat-field corrections were applied using the corresponding images from the Gemini Observatory Archive (GOA\footnote{https://archive.gemini.edu/}) and the Gemini routines included in the \textsc{IRAF} package, which were also employed for the usual processing of the data. 

The images in the $i'$ filter showed a significant fringing pattern. This effect, if left uncorrected, can affect negatively the photometric quality of the images \citep[e.g.][] {howell_2012}. 
For its treatment, images of a blank field were reduced in the same manner as the science field. The final fringe pattern was created with the task \textrm{GIFRINGE} using a combination of these dithered images, in order to dispose of any stars that may be present in them. The task \textrm{GIRMFRINGE} scales the image properly and subtracts it from the science image, thus eliminating the interference pattern. 

For each filter, the images are slightly dithered so that their combination in addition to a bad pixel mask, gives
a final product clean of cosmic rays and flaws generated by the detector. To perform a more exhaustive search for stellar-like objects, we eliminated most of the galaxy light applying a median filtering with the task \textrm{FMEDIAN} and subtracting it from the original image. We performed this task twice, using first a $200 \times 200$ pixels window and then a $40 \times 40$ pixels window, respectively. 

In addition to our GMOS-S images, a smaller WFPC2 field which contains the galaxy was obtained from the HST Data Archive. The exposure times were $4200\,s$ in both filters, $F555W$ and $F814W$. In this case, a smooth model of the galaxy's light distribution was generated using the \textmd{ELLIPSE} task from the \textmd{DAOPHOT} package in \textsc{IRAF} and subtracted from the image. These observations were carried out in May, 1999 (Proposal 6587, PI: Richstone, D.).

\section{GMOS Photometry}
\subsection{Point-source selection}
\label{sec:pss}
As GCs will appear as unresolved point-like objects at the distance of NGC\,6876, we used SE\textsc{xtractor} \citep{Bertin_1996} to detect the GC candidates. This code offers a variety of filters to improve the detection of faint objects. In this work, we combined the selection obtained with a \textit{gaussian} filter and a \textit{mexhat} filter. The latter is usually a better fit for objects in crowded areas, such as the regions closer to the galactic centre. 

Besides producing a catalog with estimated aperture photometry for each object, the software assigns a value between 0 and 1 to every object depending on how point-like they are. Those nearer to 0 are extended objects, likely to be background galaxies. The parameter which stores this value is called \textit{CLASS STAR}. Our first requirement for point-like objects was \textit{CLASS STAR} $> 0.5$.

Using tasks in the \textrm{DAOPHOT} package, we obtained approximated magnitudes of the selected objects through aperture photometry. Using these as initial values, more precise magnitudes were calculated using PSF photometry. Based on a selection of 100 bright stars distributed over the field, we obtained a second-order variable model for the PSF suitable for our images. \textrm{ALLSTAR} uses this PSF to measure the instrumental magnitude for every selected object. 

As it performs a fit for every object, it also gives the results of the goodness fits, which were used to make a refined selection, choosing the values for $\chi^2$ using the artificial stars added for completeness as described in Section \ref{section:complete}.

Our final point-sources catalog was made of $\sim 1670$ objects, in a magnitude range of $18.7 < i'_{0} < 25.5$.

\subsection{Photometric calibration}
To calibrate our instrumental magnitudes, we worked with observations of the field 195940-595000 of standard stars from \cite{Smith_2002}, taken with the science data. Given its right ascension and altitude, this field is observable at the same time as NGC\,6876, allowing the observations to be conducted under conditions as similar as possible to the science ones. Different exposure times were used to obtain images of the standards field, so that we did not lose neither too faint nor too bright objects. With the task \textrm{PHOT} we performed photometry over them, varying the aperture over a large range of sizes. 

These results were used to build a growth curve, which tells us the size of aperture needed to capture all the light corresponding to the object. In this case, the curve becomes stable at around 30\,pixels. This is the aperture used both for the final photometry of the standard stars, and for the aperture correction applied to the magnitudes previously obtained for the point-source objects in our science field.

The Gemini telescope website\footnote{http://www.gemini.edu/sciops/instruments/gmos/calibration}  provides users with the necessary transformation equations as seen in Eq.\ref{eq:std}, for which we determined the corresponding photometric zero points for each band and the colour terms shown in Table \ref{tab:std} by interpolating the relation between our instrumental magnitudes and the standard ones, obtained from the catalog in \cite{Smith_2002}. 
\begin{equation}
\begin{split}
m_{std}= & m_{zero}-2.5\,log_{10}\left(N(e-)/exptime\right) \\
& -k_{CP}(airmass-1.0)+CT(color_{1}-color_{2})
\end{split}
\label{eq:std}
\end{equation}

\begin{table}
\centering
 \caption{Coefficients of the standard transformations.}
 \label{tab:std}
 \begin{tabular}{|l|l|l|}
  \hline
  Filter & Zero point & Colour Term \\
  \hline
  $g'$ & $3.31\pm0.04$ & $0.13\pm0.06$ ($g'$-$r'$)\\[2pt] 
  $r'$ & $3.38\pm0.02$ & $0.07\pm0.03$ ($g'$-$r'$)\\[2pt]
  $i'$ & $2.98\pm0.03$ & $0.01\pm0.04$ ($r'$-$i'$)\\[2pt]
  \hline
  	\hline
\end{tabular}
 
\end{table}

Finally, we applied a correction for galactic extinction, using the absorption coefficients provided by \cite{schlafly11} with $E(B-V)=0.04$, thus obtaining the final standard magnitudes for our GC candidates.

\subsection{Completeness and background estimation}
\label{section:complete}
The estimation of the completeness of the GC candidates selection was made through the addition of artificial stars, which were then measured. In this case, 500 artificial stars were added using the task \textmd{ADDSTAR} within \textmd{DAOPHOT}, generated with random x, y positions scattered through the whole field, and a random magnitude within the usual limits for GC selection. This process was repeated 40 times, generating new stars every time, for each filter. Afterwards, the reduction process used for the science images was applied to the 120 frames, filtering them and carrying out the detection of point-sources and their photometry in exactly the same way as it was done before. No systematic trends were detected when analyzing the differences between the input and output magnitudes in relation to other parameters, such as distance to the galaxy centre.

\begin{figure}
 \includegraphics[width=0.9\columnwidth]{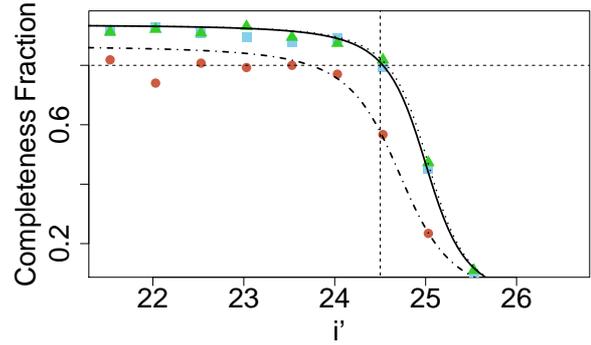}

 \caption{Completeness fraction for the inner region (dots), medium region (squares) and external region (triangles). Each region has its corresponding fit plotted in dot-dashed, solid and dotted, respectively. The dashed line shows the chosen limit.
 }
 \label{fig:compl}
\end{figure}

In order to ensure a completeness of 80\%, we take a faint  brightness limit of $i' = 24.5$. In addition, we fit an analytic function to our completeness curves, as can be seen in Figure \ref{fig:compl}, of the form:

\begin{equation}
f(m) = \beta \left( 1 - \frac{\alpha(m-m_0)}{\sqrt{1+\alpha^2(m-m_0)^2}}\right) 
\end{equation}

This function has $\beta$, $\alpha$ and $m_0$ as free parameters, and it is similar to the one used by \cite{harris2009}. It allows us to have a more exact determination of the completeness level at different galactocentric distances for when more precise corrections are necessary.

The contamination level caused by background galaxies and nearby stars was estimated through the analysis of a comparison field, also observed with Gemini South in GN-2001B-SV-67 programme. However, the correction was deemed negligible, since only around 8-10 objects were found within the colour range corresponding to GCs.

\subsection{GC selection}

\begin{figure}
 \includegraphics[width=\columnwidth]{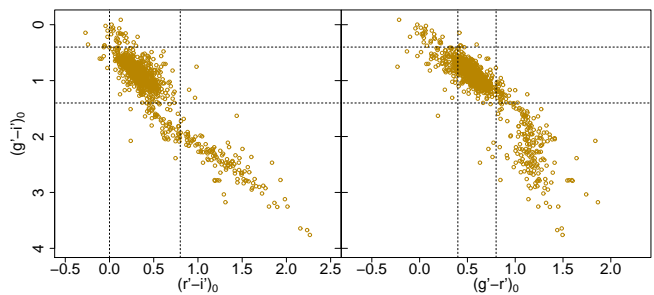}

 \caption{Colour-colour diagram for all the point-source objects detected with SE\textsc{xtractor}. Dashed lines show the limits taken for the GC selection. 
 }
 \label{fig:dcc}
\end{figure}

Since GCs usually fall within clearly defined colour ranges, we set our selection using limits derived from those used in other works with the same photometric system \citep[e.g.][]{faifer11,bassino2017}. As can be seen in the colour-colour diagrams in Figure \ref{fig:dcc}, the chosen limits are $0.0<(r'-i')<0.8$, $0.4<(g'-i')<1.4$ and $0.4<(g'-r')<0.8$. Regarding brightness limits, ultra-compact dwarfs (UCDs) are known to be present in this same colour range but they are brighter than GCs, so we apply an upper limit in magnitude derived from \cite{mieske06} to separate them from our GC candidates. For setting the faint magnitude limit, we will consider the 80\% completeness level at $i'=24.5$.

\begin{figure}
 \includegraphics[width=\columnwidth]{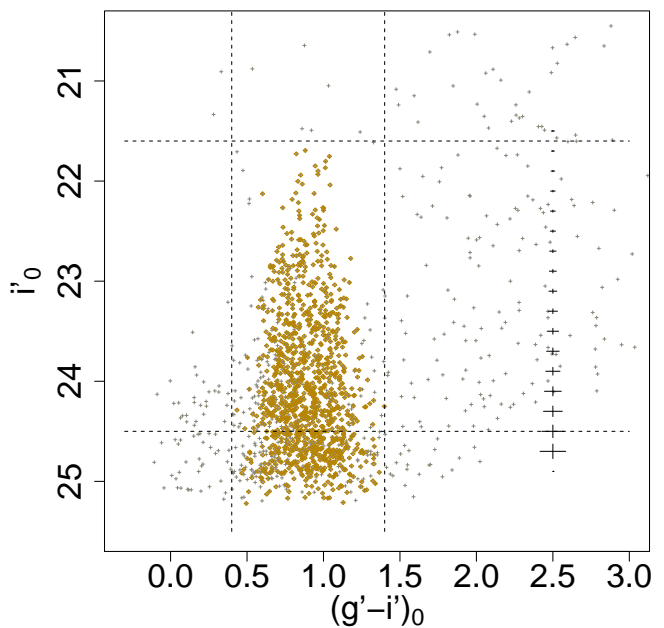}
 \caption{Colour-magnitude diagram for all the point-sources detected with SE\textsc{xtractor} (grey crosses). The golden diamonds indicate the GC candidates, selected through the combination of limits in all colours as mentioned in the text. In magnitude, the faint limit corresponds to a completeness level of 80\%, whereas the bright limit separates possible UCD candidates. On the right, the crosses represent the errors in colour and magnitude.}
 \label{fig:dcm}
\end{figure}

Figure \ref{fig:dcm} shows the colour-magnitude diagram of all the point-sources, and the $\sim 760$ GC candidates as golden diamonds. The errors depicted with horizontal lines on the right-hand side of the figure correspond to the average dispersion within each bin of 0.1\,magnitude.

In the brighter end of the GC selection, it is noteworthy how there appears to be a significant lack of GCs on the blue side. This effect, usually referred to as the `blue-tilt', has been recurrently interpreted as a mass-metallicity relation (MMR) \citep{bluetilt,usher2015,harris2017}. Though many reasons for this shift in the MMR were proposed, recent analysis performed using the E-MOSAICS simulations have suggested the effect is caused by a lack of massive blue GCs. This absence is clear in our diagram, and it is proposed to be the consequence of a physical upper limit in cluster formation \citep{usher2018}.

\begin{figure}
    \centering
    \includegraphics[width=\columnwidth]{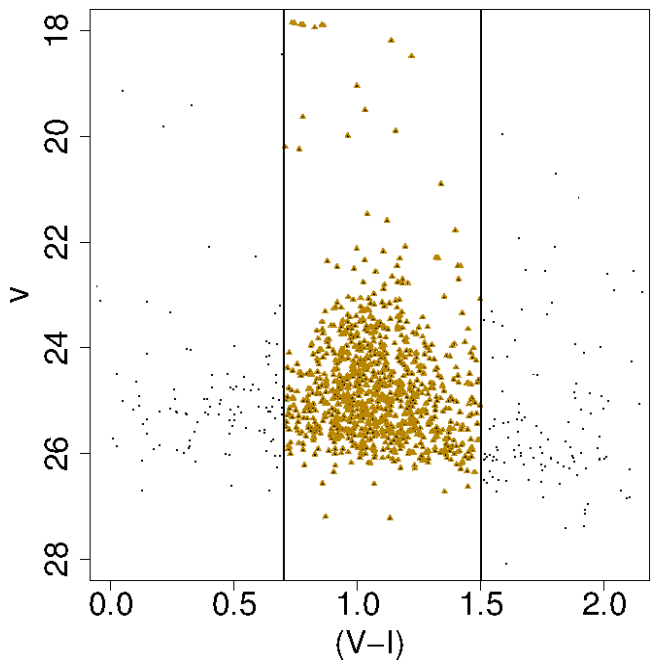}
    \caption{Colour-magnitude diagram for the point-source objects detected on the WFPC2 field (circles). The triangles indicate those that fulfill the colour criteria described in the text to be considered GC candidates.}
    \label{fig:dcmhla}
\end{figure}

In the case of WFPC2 images, the point-source selection was performed over the filtered images of the field using the same method as described above. In order to select GC candidates, colour limits were applied in the colour-magnitude diagram shown in Figure \ref{fig:dcmhla}. The ($g'$,$i'$) to ($V$,$I$) transformation was obtained from \cite{bassino2017}, obtaining colour limits of $0.7<(V-I)<1.5$. 

\begin{figure}
 \includegraphics[width=\columnwidth]{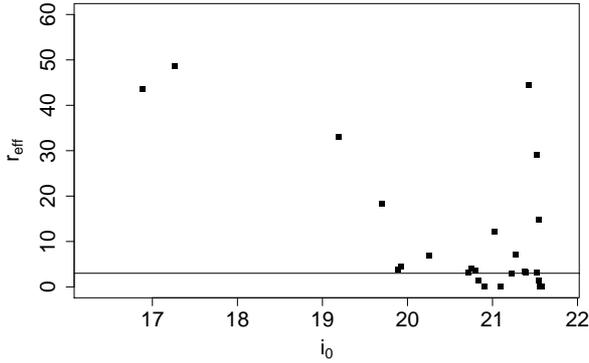}
 \caption{Distribution of the effective radius of all point-like sources with magnitudes brighter than the assumed limit for UCD candidates, against the GMOS $i'_{0}$ magnitude. The solid line indicates $\rm{r}_{\rm{eff}}=3\,pc$, the typical value for GCs \citep{jordan2007}.}
 \label{fig:radios}
\end{figure}

In addition, these images were also used to confirm the nature of the UCD candidates, previously selected only by colour and magnitude. In the case of those objects present in the field which had $i'_{0}<21.6$, ISHAPE \citep{larsen1999} was used in order to obtain estimations of their radii through the fitting of King profiles \citep{king62,king66}, with a concentration parameter of $c=30$ \citep{madrid10}. The results are shown in Figure \ref{fig:radios}. There is an accumulation near $\rm{r}_{\rm eff}\sim3$\,pc, which is the typical effective radius of GCs \citep{jordan2007}, and a sample of candidates which present radii larger than 10\,pc, which is the order as the typical effective radius of small UCDs \citep{mieske08}. This indicates the presence of a UCD population with $\sim10$ members in the inner region of the galaxy. It is important to note that, given the resolution of the WFPC2 being nearly the same size as the FWHM of stellar objects, these results are only accurate enough to be considered approximations.

\section{Globular Cluster System}
\subsection{Colour distribution}

\begin{figure}
 \includegraphics[width=\columnwidth]{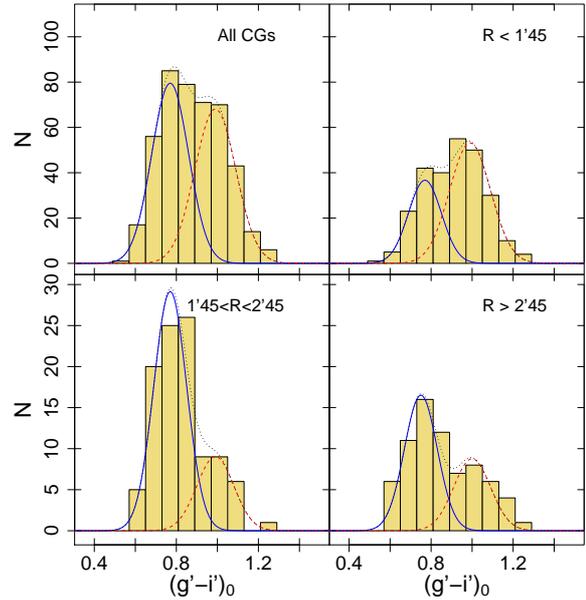}
 \caption{Colour distribution diagrams corresponding to the entire population in the top left panel, and to rings at different galactocentric distances in the following panels, moving from the inner region (top right) to the outer region (bottom right). The blue solid and the red dashed lines represent the gaussian fits corresponding to the metal-poor and to the metal-rich subpopulation respectively. The dotted black line shows the sum of both functions.}
 \label{fig:dcol}
\end{figure}

In Figure \ref{fig:dcol} we present the colour distribution for all the GC candidates in the upper left panel. The other three panels show the colour distribution for GC candidates at different galactocentric radii.
For both the complete sample and the innermost region, the minimum distance is 15\,arcsec so as to avoid the central zone, since it is saturated. 
 
In order to determinate whether our distributions are multimodal or unimodal, 
we apply the algorithm Gaussian Mixture Modeling \citep[GMM,][]{muratov2010}, which tests the sample for bimodality with two different statistics. 
The first one is the kurtosis of the distribution, which indicates how flat the distribution is. The more negative the kurtosis is, the flatter the distribution and the more likely it is to be multimodal, since an unimodal distribution is expected to have a peak. However, though a negative kurtosis is necessary for bimodality, it is not enough to confirm it.

The second statistic, D, which assumes bimodality and tells us how far apart the estimated means are, taking the dispersions into consideration as shown in Equation \ref{eq:d}. If it is greater than 2, the separation is wide enough that we can assume a bimodal distribution. 

\begin{equation}
D = \frac{|\mu_{1}-\mu_{2}|}{\left[\left(\sigma_{1}^2-\sigma_{2}^2\right)/2\right]^{1/2}}
\label{eq:d}
\end{equation}

The resulting statistics for the total GC population and three different radial ranges are shown in Table \ref{tab:gmm}. In all cases, bimodality seems to provide the best fit. It can be seen that the external region has the largest values in both tests, whereas the medium region has weaker results, which is caused by the large difference between the amount of GCs in each subpopulation. GMM also gives the estimated means and their corresponding dispersions for the fitted Gaussians, as shown in Table \ref{tab:gmm}, and those parameters are the ones used in the Gaussians shown with solid lines in Figure \ref{fig:dcol}. The blue GC candidates are identified with a blue solid line and the red ones, with a dotted red line. 

In all the histograms, the presence of the classic blue and red GCs is evident at the modal values of $\sim 0.8$ and $\sim 1.0$. While the red subpopulation dominates the sample in the inner region, the blue subpopulation surpasses it as we reach the more external regions, and it dominates the sample as a whole, as can be seen in other ETGs \citep[e.g.][]{bassino2006,usher2013}.

The separation between both samples is considered to be at $(g'-i')=0.9$, which is the colour obtained from GMM for which an object has an equal probability of belonging to either subpopulation. 

\begin{table*}
 \caption{Means, standard deviations and statistics obtained with GMM.}
 \begin{tabular}{|ccccccc|}
  \hline
\multirow{2}{*}{Region} & \multicolumn{2}{c|}{Blue Subpop.} & \multicolumn{2}{c|}{Red Subpop.} & Kurtosis
 & D \\

 & $\mu$ & $\sigma$ & $\mu$ & $\sigma$ & &  \\
  \hline
 All GCs &  $0.77\pm0.02$ & $0.09\pm0.01$ & $0.99\pm0.02$ & $0.10\pm0.01$ & -0.67 &  $2.39\pm0.24$ \\[2pt]
  \hline
Inner & $0.77\pm0.04$ & $0.08\pm0.02$ & $0.99\pm0.03$ & $0.09\pm0.02$ & -0.51 & $2.35\pm0.39$ \\[2pt]

  \hline
Medium & $0.77\pm0.04$ & $0.08\pm0.02$ & $1.01\pm0.08$ & $0.09\pm0.04$ & 0.20 & $2.68\pm0.95$\\[2pt]

  \hline
External & $0.75\pm0.02$  & $ 0.08\pm0.01$ & $1.03\pm0.04$ & $0.09\pm0.02$ & -0.91 &  $3.32\pm0.59$\\[2pt]

  \hline

\end{tabular}
 \label{tab:gmm}
\end{table*}

\subsection{Projected spatial and radial distribution} \label{radial}
\begin{figure}
 \includegraphics[width=\columnwidth]{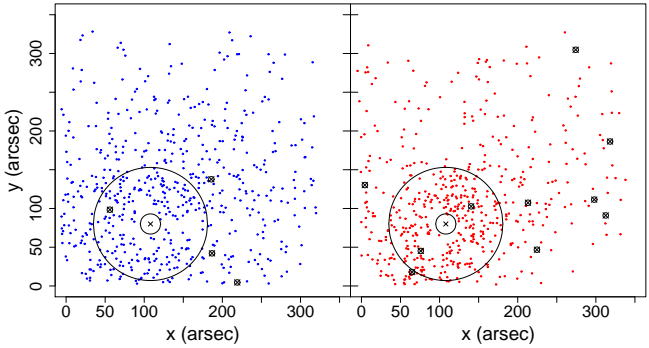}

 \caption{Projected spatial distributions for the blue (left panel) and red (right panel) GC candidates. The circles signal the region used for the azimuthal analysis in Section\,4.3. The crossed-out circles depict the location of UCD candidates.
 }
 \label{fig:espa}
\end{figure}

\begin{figure}
    \centering
    \includegraphics[width=0.75\columnwidth]{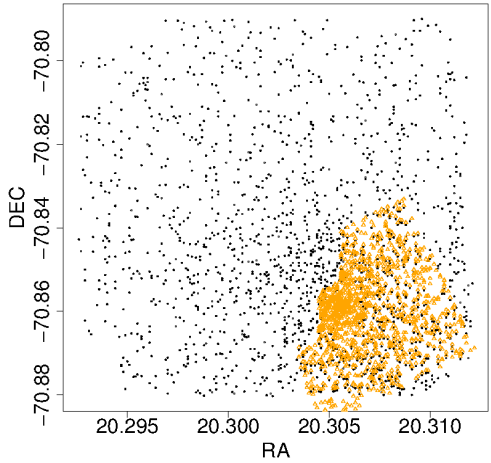}
    \caption{Spatial distribution of the GMOS sources (dots) and the WFPC2 sources (triangles), obtained with SExtractor.}
    \label{fig:esphla}
\end{figure}

The projected spatial distribution for both blue and red GC candidates in our GMOS field is presented in Figure \ref{fig:espa}. The centre of NGC 6876 is indicated with a cross. It can be seen that even on the borders of the field that are furthest from the galaxy, the density of GCs is fairly high, indicating the system is larger than our field of view (FOV). The red GCs are more concentrated towards the galaxy, whereas the blue GCs extend throughout the entire field, dominating at larger radii. Since the central region was excluded from all analyses due to the incompleteness caused by saturation, this section was complemented with the WFPC2 photometry which allows us to examine the distributions in more detail. In Figure \ref{fig:esphla}, we show the spatial distribution of the sources obtained using these images, superimposed on our GMOS distribution in order to show the region covered by them.

In order to eliminate any possible contamination due to the neighbour galaxy, we considered the variations in density throughout the field, and found they were negligible. 
Considering the intrinsinc magnitude of the galaxy, we can 
assume it does not present an amount of GCs significant enough that it might affect our results.

\begin{figure}
 \includegraphics[width=\columnwidth]{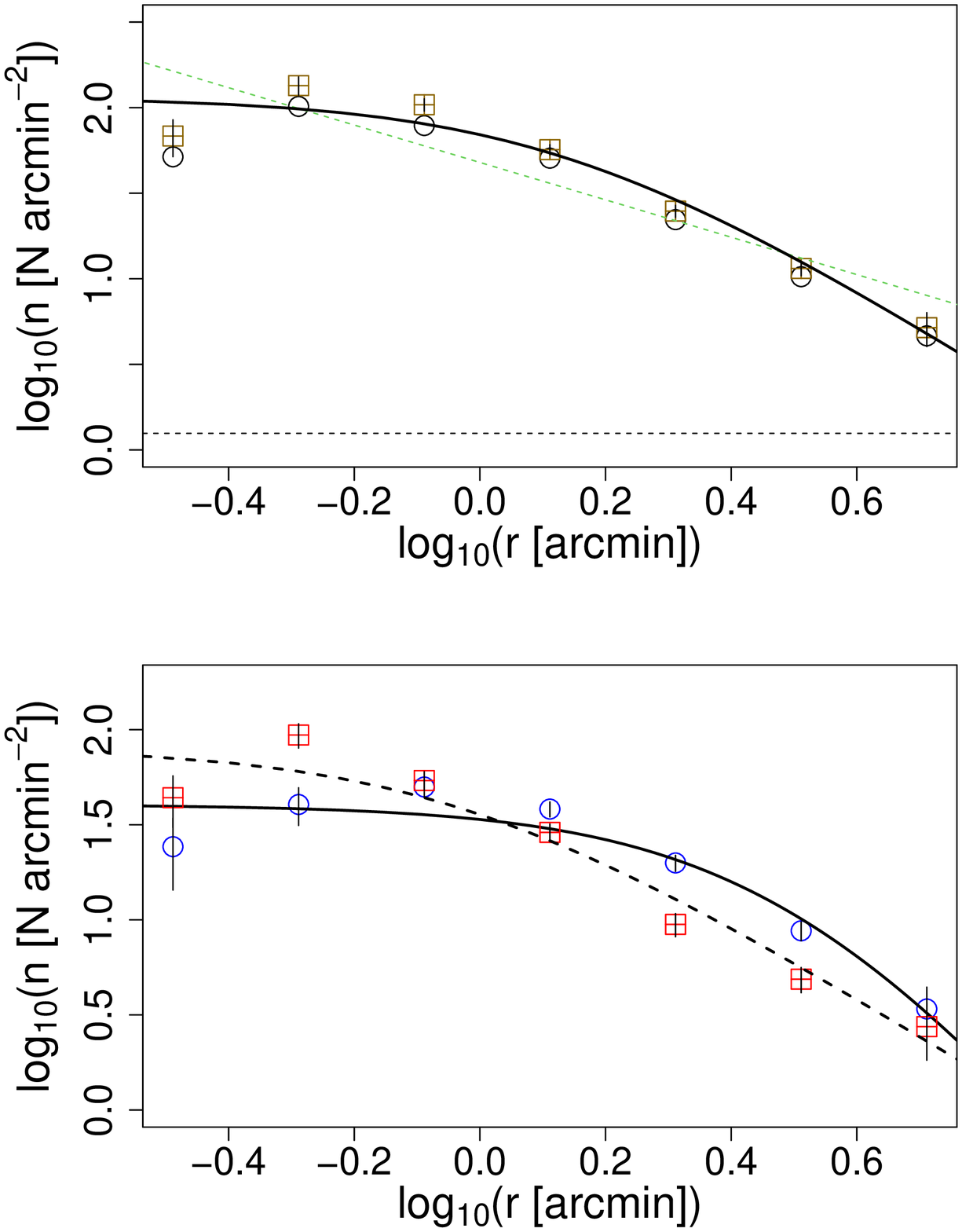}
 \caption{Projected radial distribution for the entire GC population  (top panel), both corrected for completeness (squares) and without applying the corrections (circles), and for the red (squares) and blue (circles) sub-population (lower panel). For the entire population, the solid line shows a Hubble function fit, while the dashed line shows a power-law fit. In the case of the sub-populations, they both have Hubble function fits, the solid line depicts the corresponding to the blue GCs and the dashed line, the one for the red GCs. The dotted line depicts the background level.}
 \label{fig:drad}
\end{figure}

Figure \ref{fig:drad} shows the radial distribution for the entire population in the upper panel, and for both subpopulations in the lower panel. In both cases, the sample has been corrected by completeness using the function previously fitted to the completeness curve. Power-laws were fitted to the three samples, with slopes of $-1.11 \pm0.21 $ for the entire population, $-1.25 \pm0.20 $ for the red sample and $-0.78 \pm0.25 $ for the blue sample.

These fits reinforce our previous result, showing that the red GCs decay at larger radii much faster than the blue ones.

It can be seen that all three radial profiles flatten towards the inner regions of the galaxy. Since the completeness correction was performed considering the variations in distance, this flattening might be a sign of a real lack of GCs in the vicinities of the galactic centre, rather than it just being an observational effect. In order to take the change in the slope into consideration, a modified version of the Hubble-Reynolds law (Eq.\ref{eq:hubble}, \citealt{binney1987,dirsch2003}) was fitted to the total population. In this function, $n(r)$ represents the surface number density, and $r_{0}$, a core radius. It converges to a power-law of index $2\rm{b}$ at large radii. 

\begin{equation}
\label{eq:hubble}
n(r)=\rm{a} \left(1+\left(\frac{r}{r_{0}}\right)^2\right)^{-\rm{b}}
\end{equation}
The coefficients of the best fits are indicated in Table \ref{tab:radfits}. 

\begin{table}
 \caption{Coefficients 
of the Hubble-Reynolds law fitted to the GMOS radial distributions.}
 \label{tab:radfits}
 \begin{tabular}{|llll|}
  \hline
 Population & a & $r_{0}$ & b \\
  \hline
 All GCs & $2.08\pm0.09$ & $1.29\pm0.55$ & $1.13\pm0.32$\\[2pt]
  \hline
  Red GCs & $2.22\pm0.38$& $0.39\pm0.29$ & $0.86\pm0.12$\\[2pt]
  \hline
  Blue GCs & $1.70\pm0.08$& $2.37\pm1.11$ & $1.57\pm0.69$\\[2pt]
	\hline
\end{tabular}
\end{table}

\begin{figure}
    \centering
    \includegraphics[width=0.9\columnwidth]{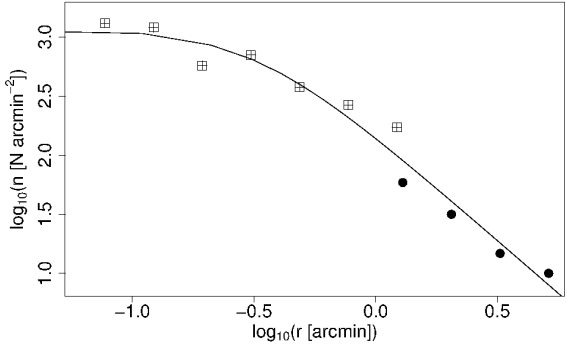}
    \caption{Radial distribution for the GMOS sources (dots) and the WFPC2 sources (squares). The solid line indicates the Hubble-Reynolds law fit.}
    \label{fig:dradhla}
\end{figure}

Since the WFPC2 sources cover the inner region of the galaxy, we combined their radial distribution with the one obtained for our GMOS sources in Figure \ref{fig:dradhla}, thus obtaining a distribution less affected by saturation, and we fitted a Hubble-Reynolds law as described in Equation\,\ref{eq:hubble}. The resulting parameters of this fit were $\rm{a}=3.1\pm0.1$, $\rm{b}=0.9\pm0.1$ and $r_{0}=0.3\pm0.1$.


The difference between this result and the one obtained merely for the GMOS sources indicates that a significant portion of the flattening in the inner region shown in Figure \ref{fig:drad} is a consequence of the difference in the noise levels, which is larger in the GMOS observations near the centre of the galaxy. However, this new radial profile still presents signs of flattening, though at a slower pace, indicating the disruption of GCs near the centre of the galaxy is significant.

Though our FOV does not reach background levels, we can extrapolate the Hubble-Reynolds fits to obtain an estimation of the full extension of the GCS. For this, we assume the background level to be at N/${\rm arcmin}^2 \sim 1.25$, as obtained for similar works \citep{caso2015,bassino2017}. The estimated extension is $\rm{r}_{\rm{GCS}}\sim 125$\,kpc. Through integrating the Hubble-Reynolds law until that point, we can also obtain an approximate number for the total population of the GCS within this extension and brighter than $i'_{0}=24.5$ which gives us $\sim 1960$ GCs.

\subsection{Projected azimuthal distribution}    
\begin{figure}
 \includegraphics[width=\columnwidth]{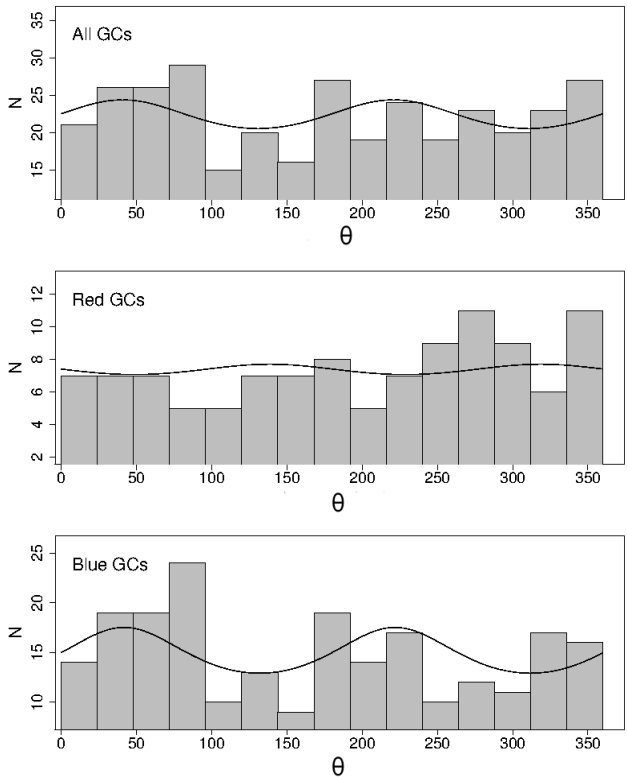}
 \caption{Projected azimuthal distribution for the entire GC population (top panel) and for the red (middle panel) and blue (circles) GC subpopulations (lower panel). The solid lines correspond to the fits described in the text.}
 \label{fig:dacim}
\end{figure}

In order to achieve azimuthal completeness, we decided to only consider GCs with galactocentric radii between 13,6 and 75\,arcsec, as shown in Figure \ref{fig:espa}. The projected azimuthal distribution is shown in Figure \ref{fig:dacim}.

We studied this distribution using the expression from \cite{mclaughlin1994}:

\begin{equation}
\sigma(R, \theta) = k R^{-\alpha}[cos^2(\theta-PA)+(1-\epsilon)^{-2}sin^2(\theta-PA)]^{-\alpha/2}
\end{equation}

where $\sigma$ is the number of GC candidates, and a power-law radial distribution is assumed, with $\alpha$ being the value obtained in Section \ref{radial}. The position angle (PA) is measured counterclockwise from the North and fitted alongside the ellipticity. The number of GC candidates $\sigma$ is calculated for circular regions of 24 degrees, as shown in Figure\,\ref{fig:dacim}. 

\begin{table}
 \caption{Values of position angle and ellipticity obtained for the complete sample and for the two subpopulations (those for the red subpopulation are not significant).}
 \centering
 \begin{tabular}{|llll|}
  \hline
 Parameter & Total & Red GCs & Blue GCs \\
  \hline
 PA & $40.7^{\rm{o}}\pm23^{\rm{o}} $ & $48.9^{\rm{o}}\pm7.8^{\rm{o}}$& $42^{\rm{o}}\pm8.6^{\rm{o}}$\\[2pt]
  \hline
 $\epsilon$ & $0.14\pm0.10$& $0.01\pm0.15$ & $0.24\pm0.11$\\[2pt]
	\hline
\end{tabular}
\label{tab:acimfits} 
\end{table}

In Table\,\ref{tab:acimfits} we present the results of the fit applied to the complete GC sample and to each of the subpopulations. It can be seen that the red GC subpopulation shows no signs of elongation, while the blue subpopulation has a much more significant ellipticity. The behavior of the complete sample seems to be led by the blue subpopulation. 

An interesting though subtle feature in the red GC subpopulation is the slight overdensity between 250\,degrees and 360\,degrees (Fig\,\ref{fig:dacim}). This happens to be in the same direction as the X-ray trail detected by \cite{machacek2005}, and the coincidence is even more clear in comparison with Figure\,7 in \cite{machacek2009}. This effect only being noticeable in the red subpopulation could be a consequence of the size of region taken into account for the azimuthal distribution. Since it is very close to the galactic centre, the ring contains an important fraction of red GCs but a less relevant fraction of blue ones, the latter being therefore a noisier sample. There is no evidence of this implied disturbance in the stellar component of the galaxy, but this is not entirely unexpected. As said by \cite{alamo2017}, numerical simulations show that dark matter halos are more affected than stars by gravitational interactions between galaxies, and among the stellar population, GCs are usually the first to be affected by the stripping because the GC system is spatially more extended than field stars \citep{smith2013,mistani2016}.

\subsection{Luminosity function and GC population}

\begin{figure} 

 \includegraphics[width=\columnwidth]{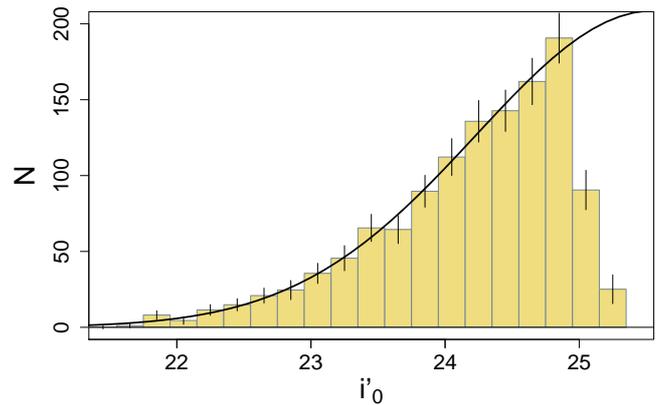}
 \caption{Luminosity function of the GC candidates. The vertical bars symbolize the error corresponding to each bin, and the solid line represents the Gaussian fit.}
 \label{fig:lumfun}
\end{figure}

Fig.\ref{fig:lumfun} shows the luminosity function of the system. One of the most striking characteristics of GCSs is the apparent universality of the luminosity function, which has a nearly identical shape in most galaxies despite the variety of formation histories and tidal effects \citep{harris2014}. Though it does depend slightly on galaxy mass \citep[e.g.][]{villegas2010}, the peak of the Gaussian function usually selected to perform the fits (\textit{turn-over}) is universal enough that it has been used as a standard candle for measuring extragalactic distances. It has been estimated that in ETGs, the turn-over of the GC luminosity function in the $V$-band is $M_{\rm V}\sim-7.4$ \citep[e.g.][]{richtler2003,jordan2007}. 

Though it can be seen from Fig.\ref{fig:lumfun} that our $i'_{0}$ images are not deep enough to reach the turn-over, we can estimate it through fitting a Gaussian to the distribution. The estimated turn-over is found at $i'_{0}=25.6\pm0.4$, which translates to a distance modulus of $(m-M)=33.5\pm0.4$ using Equation 2 in \cite{faifer11} which allows us to obtain $V_{0}$ assuming a mean value for the GCs with $i'_{0}>25.5$ of $(g'-i')_{0}\approx0.9$.

Though the dispersion in the numerous distance measurements that can be found in NED for this galaxy is large, this value is a good approximation to the one measured by \cite{blakeslee2001} using surface brightness fluctuations (SBF). 

Integrating the luminosity function allows us to estimate what percentage of the total GC population is covered up until the limit established by the completeness level. In this case, the integration revealed our $\approx 2000$ GC candidates are barely $\approx 21\%$ of the complete population, which gives us a total of $\approx 9500\pm2500$ GCs.

Another parameter usually analyzed when studying GCSs is their \textit{specific frequency}, defined simply as $S_{N}=N_{GC}\, 10^{0.4(M_{V}+15)}$ \citep{harris1981}. Thought to be almost constant among ETGs at first, it is now accepted that it varies with luminosity, and is linked to the efficiency in GC formation and retention. The ``average'' value for elliptical galaxies was clasically thought to be $S_{N}\approx3.5$. Currently, the range has extended up to 10 and even higher, showing a significant dependence with the brightness of the galaxy \citep{harris2013}. For NGC\,6876, we obtain a value of $S_{N}\approx7.74$, which falls within the range though it is high for the estimated absolute visual magnitude we considered for it.

\section{Surface brightness profile of NGC~6876}

\begin{figure}
 \includegraphics[width=\columnwidth]{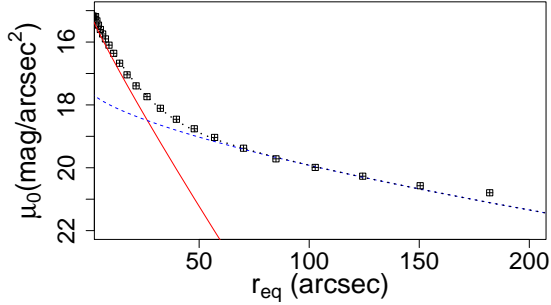}
 \caption{Surface brightness profile of NGC\,6876 in $i'_{0}$. The solid line and the dashed line represent the inner and outer components, respectively, while the dotted line represents the sum of both.}
 \label{fig:gal}
\end{figure}

\begin{figure}
 \includegraphics[width=0.9\columnwidth]{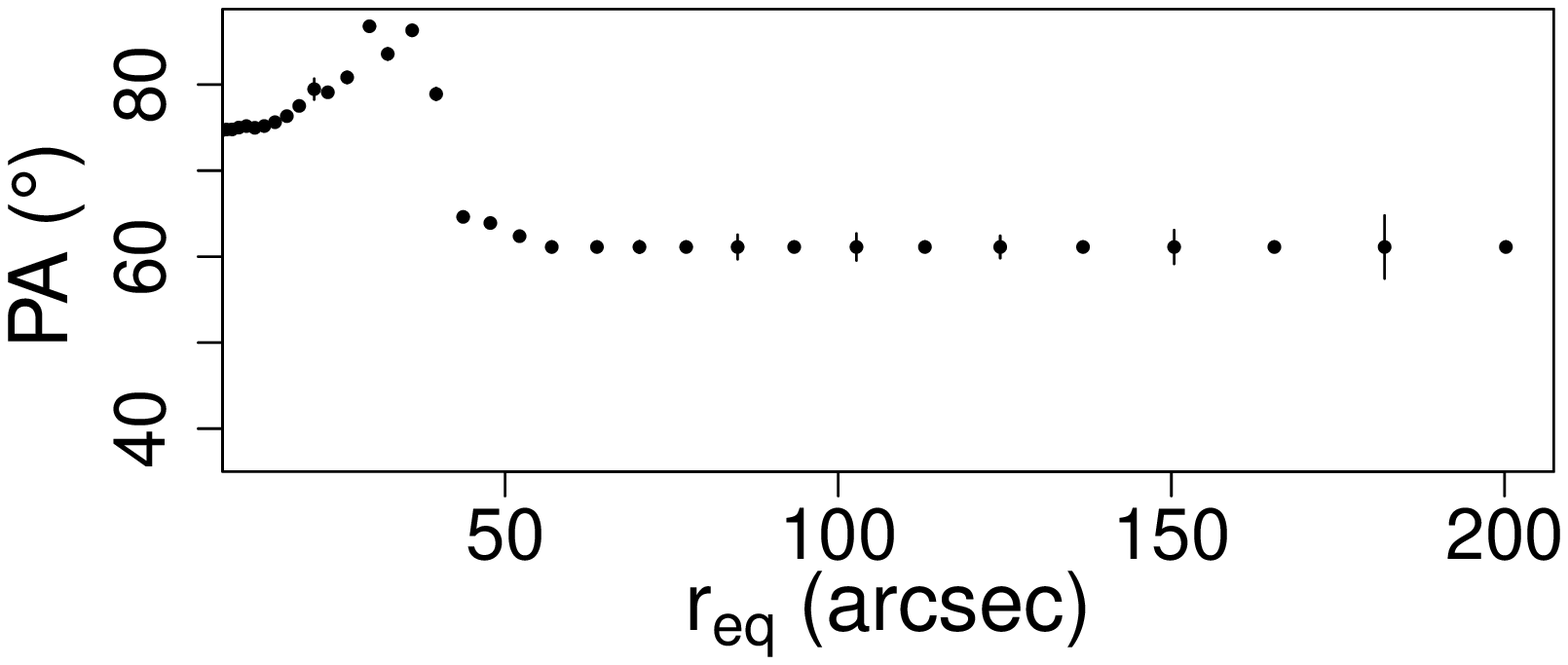}
 \includegraphics[width=0.9\columnwidth]{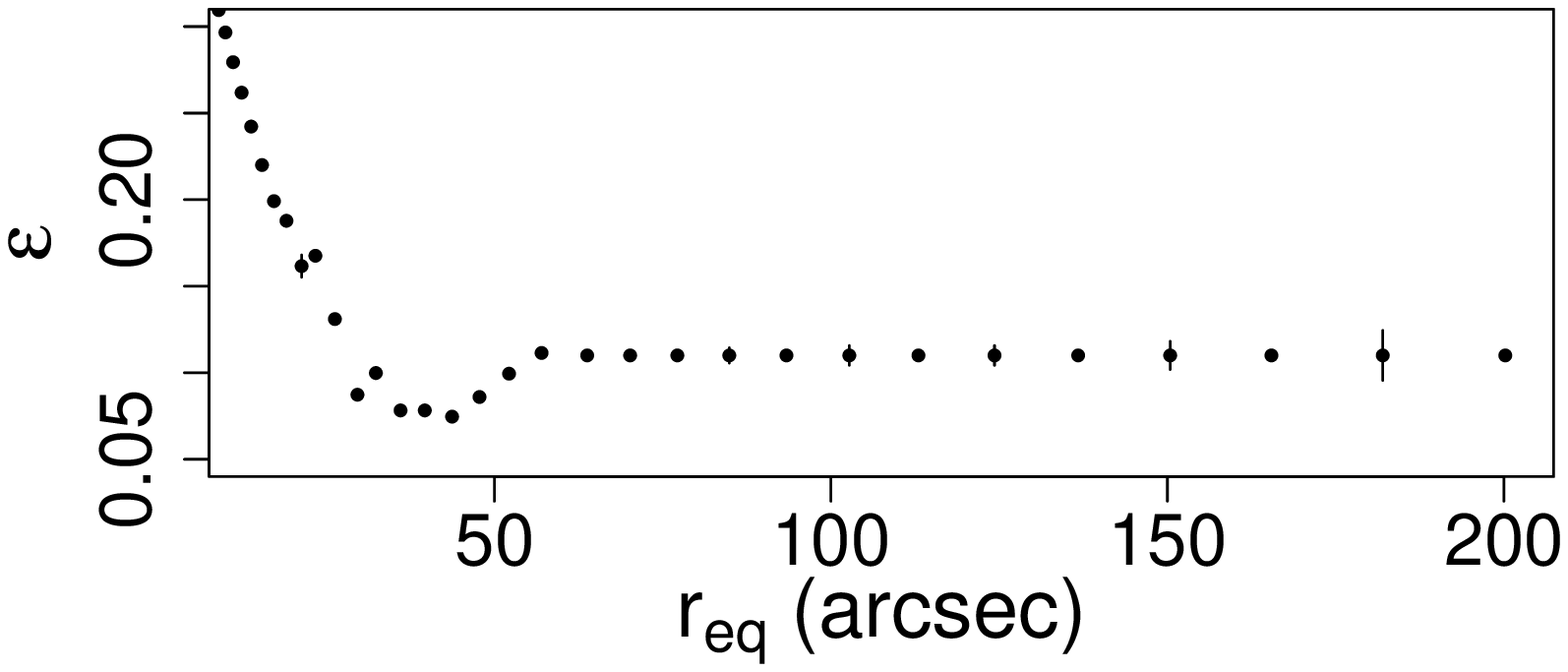}
 \includegraphics[width=0.9\columnwidth]{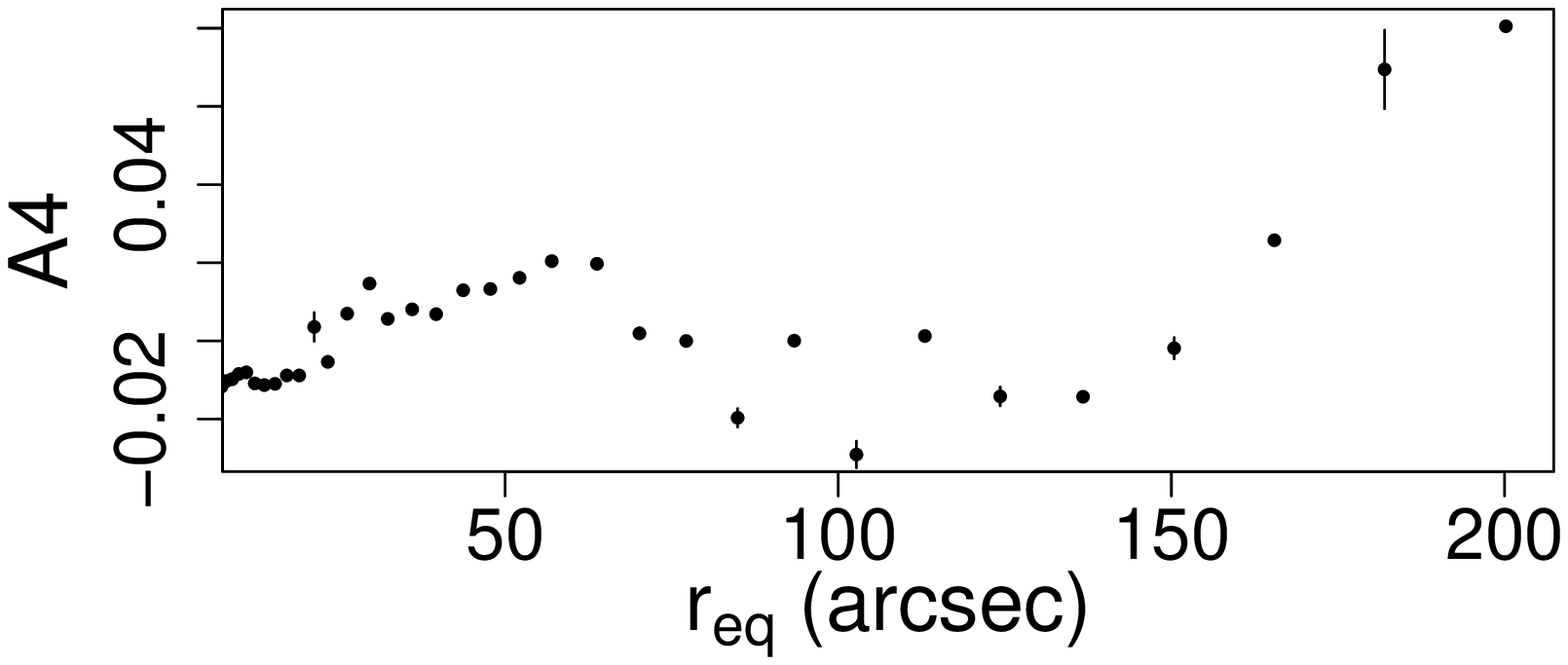}
 \caption{Upper panel: position angle of the elliptical isophotes versus effective radius. Middle panel: ellipticity of the elliptical isophotes versus effective radius. Lower panel: Parameter $A_{4}$ versus effective radius.}
 \label{fig:pa}
\end{figure}

Using the \textmd{ELLIPSE} task from the \textmd{DAOPHOT} package in \textsc{IRAF}, we studied the surface brightness profile of the host galaxy NGC\,6876 in the $i'$ filter. Other bright galaxies present in the field were masked with the same task. Within a galactocentric radius of 1\,arcmin, the isophotes were fitted leaving all parameters free, which allows us to analyse the behavior of the ellipticity and the position angle, as depicted in Figure \ref{fig:pa}. For isophotes at larger radii, those parameters were set fixed to the last values obtained, since the fact that they have regions that fall off the observed field, makes it impossible to get a proper fit. The surface brightness, however, can still be modeled accurately enough as shown in Figure \ref{fig:gal},  which shows the surface brightness versus equivalent radius ($r_{eq}=\sqrt{ab}=a\sqrt{1-\epsilon}$ where $a$ is the isophote semi-major axis and $\epsilon$ is the ellipticity). Two S\'ersic models were fitted \citep{sersic1968}.  

\begin{equation*}
\mu(r)=\mu_{0}+1.0857\left(\frac{r_{eq}}{r_{0}}\right)^{1/n},
\end{equation*}

where $\mu$ is expressed in mag\,$\rm{arcsec}^{-2}$, $\mu_{0}$ is the central surface brightness, $r_{0}$ is a scale parameter and $n$ is the S\'ersic shape index. The resulting parameters for the inner and outer components are presented in Table\,\ref{tab:gal}. They point to the existence of an inner disc and an outer spheroid.

\begin{table}
 \caption{Parameters of the two S\'ersic models fitted to the galaxy profile in the $i'$ filter.}
 \centering
 \begin{tabular}{|llll|}
  \hline
 Parameter & $\mu_{0}$ & $r_{0}$ & $n$ \\
  \hline
 Inner & $15$ & $7.32$& $1.1$\\[2pt]
  \hline
 Outer & $17.5$& $30$ & $1.5$\\[2pt]
	\hline
\end{tabular}
\label{tab:gal} 
\end{table}

Through comparison with the surface brightness profile obtained from the Carnegie-Irvine Galaxy Survey \citep{ho2011}, it becomes evident that the background level has not been reached in our field of view. In the Atlas profile, the background level is  $\sim24\,\textrm{mag\,arcsec$^{-2}$}$ in the $I$ filter. 

The parameters of the isophotes, i.e. position angle, ellipticity and $A_{4}$ Fourier coefficient, were measured with \textmd{ELLIPSE} and are presented in Figure \ref{fig:pa}. 
The upper panel shows the position angle PA as we move further from the galaxy centre. The sudden change observed at around $r_{eq} \sim 40$\,arcsec matches the end of the drop in ellipticity $\epsilon $ at around the same radius, shown in the middle panel of Figure \ref{fig:pa}. The radius corresponding to such changes in PA and $\epsilon $ approximately agree with the galactocentric radius at which the inner component becomes fainter than  the outer one, and the total surface brightness is dominated by the outer spheroid. 
Since the isophotes turn more circular as we increase the radius, the errors in the estimation of the PA turn into very large values. 
In the case of the $A_{4}$ parameter, we see the inner regions have a negative value, which would indicate the central isophotes are boxy. Then it switches to positive after $r_{eq} \sim30$\,arcsec. Though it turns negative again at larger radii, the errors become too large to gather any conclusions.  

\section{Discussion}

Our photometric analysis of the GCS of NGC\,6876 reveals a hint of interaction with NGC\,6872 in the red subpopulation, though the rest of the results do not show further evidence.

The GMM test gives a solid bimodal fit for the colour distribution, and once separated the two GC subpopulations show normal behaviour, corresponding to what is usually find in the literature. However, the total GC  population as calculated through the integration of the luminosity function is quite high, meaning NGC\,6876 has a very populated system. Its extension is big, with a value calculated through the extrapolation of the radial distribution of about $\sim125\,kpc$. 

In order to obtain a better approximation of the full extent, we use the relation between the GCS spatial extent versus galaxy stellar mass given by \cite{kartha2014}, deriving the host stellar mass by means of the mass to light ratio given by \cite{zepf1993}. Then, we obtain $\log(M/M_{\odot})=12$ which results in an extension of $\approx 88\,Kpc$ for the GCS, which is a severe underestimation according to our results. However, this is not the only rich GCS that shows a larger extension than the one obtained from the \cite{kartha2014} correlation \citep[e.g.][]{bassino2006,blom2012,caso2017}.

We can also estimate the halo mass (i.e. baryonic plus dark mass), using the method described by \cite{harris2017b}. From the luminosity function we obtained a total population of $N_{GC}\sim9400$ GCs. Adopting a GC mean mass of $<M_{GC}>\sim3.5\times10^{4}\,M_{\odot}$, it results in a total mass for the galaxy's GCS of $M_{GCS}\sim3.25\times 10^{8} M_{\odot}$. Dividing this result by the number ratio, a constant introduced by \cite{blakeslee1997} and defined as $n_{N}=N_{GC}/M{h}$, recalibrated by \cite{harris2017} to a value of $n_{N}=2.9\times 10^{-5}\pm0.28$, we end up with a halo mass of $M_{h}\sim1.12\times 10^{13} M_{\odot}$. This mass is in agreement with other ETGs with highly populated GCSs \citep{pedersen97,nak00,caso2017}.

\begin{figure} 
\centering
 \includegraphics[width=0.7\columnwidth]{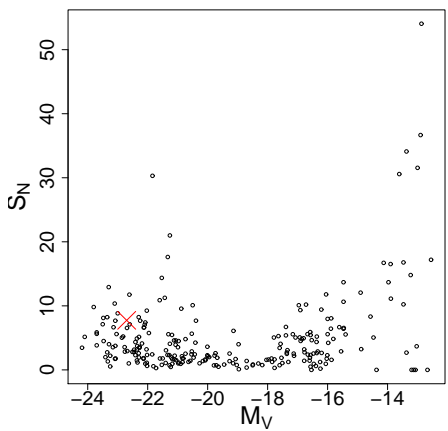}
 \caption{Specific frequency of GCSs in ETGs versus galactic absolute $V$ magnitude, 
 with NGC\,6876 represented by a red cross.}
 \label{fig:snr}
\end{figure}

\begin{figure} 
\centering
 \includegraphics[width=0.7\columnwidth]{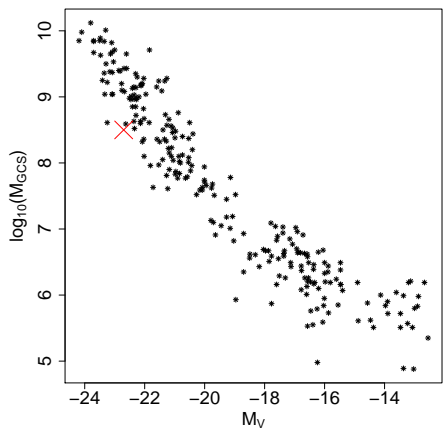}
 \caption{Mass of GCSs in ETGs versus galactic absolute $V$ magnitude, with NGC\,6876 marked with a red cross.}
 \label{fig:masscgs}
\end{figure}

Using the existing catalogue of globular systems in galaxies presented by \cite{harris2013} and selecting those in ETGs only, we compared the obtained $S_{N}$ and $M_{GCS}$ with those for similar galaxies, most of them inhabiting clusters. It can be easily seen that in both cases (see Figures \ref{fig:snr} and \ref{fig:masscgs}), the values fall among the expected ones for ETGs of this luminosity. Most ETGs in the bright end are located in dense environments, where they are expected to have undergone a rich history of mergers \citep{jimenez2011,sch14}. This explains their highly populated GCS, dominated by the metal-poor subpopulation, in the context of its build up in two phases \citep{forbes2011,caso2017}. The fact that NGC\,6876 presents both a highly populated GCS and a large stellar mass could indicate its history has also involved major mergers, despite its environment being of such low density. 

Having the total population and the stellar mass allows us to estimate another parameter, $T_{N}$, as defined by \cite{zepf1993}. This parameter is another option when looking to find a relation between GCs and the morphology and mass of their host galaxy. In our case, we obtain $T_{N}\sim9.4$, which falls among the expected values for ETGs with stellar masses similar to NGC\,6876 \citep{peng2008}.

Using the XMM-Newton observations analyzed in \cite{machacek2005}, and the expression given by \cite{grego},
we can obtain an estimation of the stellar mass through an alternate method and compare it with the one derived from the mass to light ratio. In this case, we use $\mu=0.6$ as the molecular weight, a temperature of $k T_{e}=0.76\,kE_{v}$, and we calculate the mass up to the radial limit derived from the radial projected distribution of GCs, $r=125$\,kpc. The surface brightness profile for the X ray emission was fitted using the combination of two $\beta$-models, with a core radius of $r_{c}=5$\,kpc and a value of $\beta=0.65$ for the galactic component of NGC\,6876, and $r_{c}=50$\,kpc and $\beta=0.3$ for the surrounding intra-cluster medium. Considering the combination of both profiles, a result for the stellar mass of $M\sim10^{12}\,M_{\odot}$ is achieved, which is in agreement with the one previously calculated.

One very debated scaling relation between GCSs and their host galaxies is the matter of relative size, which has been studied in comparison with both the stellar and the halo mass. Since we have estimations for both as well as for the GCS extension, it is of interest to see where our data falls in the proposed existing relations. Recently, \cite{forbes2017} suggested an approximately linear relation between GCS size and halo mass to the 1/3 power, in contraposition to the simultaneous proposal by \cite{hudson2018}, which has a much higher slope $(0.80)$. In the case of stellar mass, again \cite{hudson2018} determines a bigger slope than \cite{forbes2017}. For the halo mass, NGC\,6876 falls far from both the relation proposed by \cite{hudson2018} which gives an estimated result of $\log{\rm{r}_{\rm{GCS}}}\sim1.3$, and from the \cite{forbes2017} relation for which the size is ($\log{\rm{r}_{\rm{GCS}}}\sim1.2$). However, the errors involved in the estimation of the halo mass are quite large, so this result should only be considered as an approximation.

In the case of the stellar mass, NGC\,6876 seems to fall closer to \cite{forbes2017}, $\log{\rm{r}_{\rm{GCS}}}\sim1.9$, but it still results in a much smaller GCS when compared with our results, suggesting galaxies with highly populated GCSs show more massive halos and do not follow the linear relations proposed so far.

\begin{figure}
    \centering
    \includegraphics[width=0.75\columnwidth]{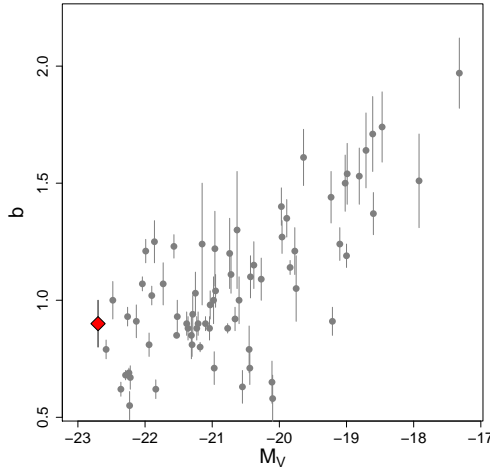}
    \caption{Parameter b as obtained from fitting a Hubble law profile to the radial distribution of ETGs, shown as black dots. NGC\,6876 is highlighted in a larger red diamond.}
    \label{fig:rads}
\end{figure}

\cite{machacek2005} fitted the X-ray emission up to 100\,kpc from the centre of the galaxy using a beta profile with two components, leading to a estimated value of $r_{500}\sim370$\,kpc. This indicates a dark matter halo that is at least that large, but looking at the profile, we can assume it extends much further. Another tracer of the dark matter halo is the distribution of blue GCs \citep{bassino2006,usher2013} since they are thought to map out the total mass of the galaxy. When we compare the b parameter in our radial distribution to the one corresponding to bright ETGs in the centre of clusters (Figure \ref{fig:rads}), the value is similar to those with approximately the same magnitude as NGC\,6876. This reinforces the suggestion that, despite being in a low density enviroment, the dark matter halo is similarly large to the halos found in galaxies in much more dense habitats.

Despite having  only a suggestion of a connection between NGC\,6876 and the trail connecting it to its spiral neighbor NGC\,6872, in the projected spatial distribution of the GCs, we can still draw some conclusions about the evolutionary history of the galaxy. The radial extent of the GCS indicates the presence of a massive halo with a significantly large extension \citep{hudson2014}. As discussed in the Introduction, in low density environments elliptical galaxies are thought to have increased their mass after $z\sim2$ through the so-called ``dry'' mergers. These involve satellite galaxies that, by the time of this interaction, are already red and quiescent thus providing a small amount cold gas and dust (if any) to the more massive galaxy and prompting little to no star formation \citep{huang2016}. Based on its colour, $(B-I)\sim2.24$ according to the Carnegie-Irvine Survey \citep{carnegie}, and its magnitude, NGC\,6876 is red enough that we can assume ``dry'' mergers as the main causes of growth.

From the webtool CMD\,3.0\footnote{http://stev.oapd.inaf.it/cmd}, assuming \citet{bre12} isochrones and Chabrier IMF, solar metallicity, typical of bright ellipticals, and the reddening corrections from \citep{schlafly11}, we estimate the age of the population for different colours, obtained from the Carnegie-Irvine Survey in the case of $B$, $V$, $R$ and $I$, and from 2Mass in the case of $J$ and $H$ \citep{huchra}. The estimated age is between $7-10\,\rm{Gyrs}$. This indicates that the galaxy is dominated by an old stellar population, which goes in agreement with the increase in mass since $z\sim2$ being prompted by ``dry'' mergers, with no recent stellar formation.

A summary of important characteristics of NGC\,6876 including our obtained parameters can be found in Table \ref{tab:final}.

\begin{table}
 \caption{Properties of NGC\,6876.}
 \centering
 \begin{tabular}{|ll|}
  \hline
 Parameter & Value \\
  \hline
 $M_{V}$ [mag] & $-22.7$ \\
 \hline
 Distance [Mpc] & $45$\\
 \hline
  $log(M_{\rm stellar}/M_{\odot})$ & 12\\
   \hline
  $log(M_{\rm halo}/M_{\odot})$ & 13\\
   \hline
  $E(B-V)$ & 0.04\\
     \hline
  $r_{GCS}$ [kpc] & 125\\
	\hline
	$T_{N}$ & 9.4 \\
	\hline
	$S_{N}$ & 7.74\\
	\hline
	
\end{tabular}
\label{tab:final} 
\end{table}

\section{Conclusions}

We performed a photometric study of the early-type galaxy NGC\,6876 and its globular cluster system, on the basis of Gemini-GMOS images in the bands $g'$, $r'$, and $i'$. The properties of the GCS were studied in detail, including the colour as well as radial and azimuthal projected distributions, luminosity function and determination of the total GC population and specific frequency. 
In addition, we made use of relations between characteristics of the GCS and of its host galaxy to estimate the stellar and halo (dark plus baryonic) mass, comparing the different proposed relations that are being currently discussed. 
Our conclusions are summarized in the following:

\begin{itemize}
\item The colour distribution of the system is bimodal, revealing the usual subpopulations, with mean values similar to the ones typically found in the literature. Further analysis of the behaviour of the color distribution in different galactocentric rings shows the red subpopulation decreases as we move towards the outer borders of the field.
\item In projection, the blue subpopulation is much more extended than the red one, as can be seen in both the spatial distribution and the radial one. This last one also allows us, through the fitting and integration of a Hubble profile to the distribution of the entire profile, to obtain an estimation for the population of the system within our completeness limit. From the azimuthal distribution we can see the blue subpopulation is much more elongated, whereas the red one presents a slight overdensity that cannot be overlooked. Interestingly, such overdensity is located in the same direction of the observed X-ray trail between NGC\,6876 and NGC\,6872, and could be indeed a sign  of their interaction. 
\item The observed sample does not reach the turn-over in the GC luminosity function. Nonetheless, the extrapolation of the fitted Gaussian function provides us with a value for it that is consistent with the distance estimated by \cite{blakeslee2001}. The integration of this fit also lets us obtain an estimation of the total population of the system, $\sim9500$ GC candidates.
\item The current relations proposed for GCS size and halo mass all seem to underestimate the measurements obtained extrapolating our radial distribution, which is $\sim125$\,kpc. 

\end{itemize}

\section*{Acknowledgements}

We thank the referee for a constructive report that helped to improve this paper. This work was funded with grants from Consejo Nacional de Investigaciones   
Cient\'{\i}ficas y T\'ecnicas de la Rep\'ublica Argentina, Agencia Nacional de Promoci\'on Cient\'{\i}fica y Tecnol\'ogica, and Universidad Nacional de La Plata, Argentina. \\
Based on observations obtained at the Gemini Observatory (GS-2013B-Q-37), which is operated by the Association of Universities for Research in Astronomy, Inc., under a cooperative agreement with the NSF on behalf of the Gemini partnership: the National Science Foundation (United States), the National Research Council (Canada), CONICYT (Chile), the Australian Research Council (Australia), Minist\'{e}rio da Ci\^{e}ncia, Tecnologia e Inova\c{c}\~{a}o (Brazil) and Ministerio de Ciencia, Tecnolog\'{i}a e Innovaci\'{o}n Productiva (Argentina). This research has made use of the NASA/IPAC Extragalactic Database (NED) which 
is operated by the Jet Propulsion Laboratory, California Institute of Technology, 
under contract with the National Aeronautics and Space Administration. This research has made use of the SIMBAD database, operated at CDS, Strasbourg, France.




\bibliographystyle{mnras}
\bibliography{biblio} 








\bsp	
\label{lastpage}
\end{document}